# Auditing spreadsheets: With or without a tool?

Simone Schalkwijk, Felienne Hermans, Michiel van der Ven, Hans Duits
HU University of Applied Sciences Utrecht and Delft University of Technology
Simone.Schalkwijk@student.hu.nl, f.f.j.hermans@tudelft.nl, michiel.vanderven@hu.nl, hans.duits@hu.nl
Additional research material is available on:
http://figshare.com/authors/Simone_Schalkwijk/706743

**ABSTRACT**

*Spreadsheets are known to be error-prone. Over the last decade, research has been done to determine the causes of the high rate of errors in spreadsheets. This paper examines the added value of a spreadsheet tool (PerfectXL) that visualizes spreadsheet dependencies and determines possible errors in spreadsheets by defining risk areas based on previous work. This paper will firstly discuss the most common mistakes in spreadsheets. Then we will summarize research on spreadsheet tools, focussing on the PerfectXL tool. To determine the perceptions of the usefulness of a spreadsheet tool in general and the PerfectXL tool in particular, we have shown the functionality of PerfectXL to several auditors and have also interviewed them. The results of these interviews indicate that spreadsheet tools support a more effective and efficient audit of spreadsheets; the visualization feature in particular is mentioned by the auditors as being highly supportive for their audit task, whereas the risk feature was deemed of lesser value.*

## 1 INTRODUCTION

A little while after spreadsheets were created in the 1960s, researchers started to examine the high rate of mistakes in spreadsheets [Olson & Nilsen, 1987]. Panko [1998] states that 86% of all spreadsheets contain errors. Eusprig collects 'horror stories' caused by errors in spreadsheets [EuSpRIG Horror Stories, 2012]. One of those stories concerns an understatement of six million US dollar of Knox County Trustee's Office's cash on hand due to a bad linked spreadsheet. This resulted in an additional audit fee of 12,500 US dollar. This example provides an insight in the possible outcome of an error in a spreadsheet. In this case the auditor did found the error but what happens when the error is not caught? This could cause serious image and financial damage to the auditor. Many tools have been developed to reduce this risk and to provide a more effective and efficient audit. What is the added value of such tools? This paper determines the perceptions of auditors about the added value of spreadsheet tools in financial auditing and provides guidelines for the improvement of spreadsheet tools. In order to get a clear overview of the perceptions of auditors about the usefulness of a spreadsheet tool during the audit, auditors from one of the big four audit firms in the Netherlands were shown functionalities in a spreadsheet tool called PerfectXL. The auditors were questioned by means of semi-structured interviews.

This paper is structured as follows. Section 2 provides a theoretical background on the most common errors in spreadsheets. Section 3 describes the methodology that has been followed. Section 4 describes the interview results. Section 5 states a conclusion. Section 6 provides topics for future research.

## 2 THEORETICAL BACKGROUND

### 2.1 Overview of errors

For many years spreadsheets are known to be error-prone [Panko,1998]. Researchers Powell, Baker and Lawson [2008b, 2009a, 2009b] focus their studies on errors in operational spreadsheets. Firstly they critically overlooked the existing literature on spreadsheet errors. They found that laboratory experiments provide evidence for high error rates measured by cell error rates or percent of spreadsheets with errors. Studies using field audits in general show the same as laboratory experiments but methods and results vary widely. Furthermore they examined errors in operational spreadsheets and the impact of those errors on financial performance. They found that 14 out of the 25 spreadsheets contained errors. In these spreadsheets 381 potential errors were found of which 117 were confirmed as errors by the developers of the spreadsheet. Among these confirmed errors, 47 had no quantitative impact on the results. But among the 70 confirmed errors, the largest error had an impact of 100 million US dollars.

Many scientists have researched the causes for the high error rate in spreadsheets and have defined a categorization of errors [Kreie, Cronin, Pendley, & Renwick, 2000; Panko, 2008; Panko & Aurigemma, 2010]. This categorization by scientists of the most common errors in spreadsheets differs. Powell, Baker and Lawson [2008b] discuss that there is no one single correct categorization of spreadsheet errors. We found the categorization by de Ruijter and Pjoter [2010] most suitable for our research because this categorization is derived from the viewpoint of a controller within an organization. They define seven categories:

1. Reference errors. This category includes errors like wrong references to other spreadsheet cells or incorrect summation of values.

2. Cells containing an incorrect formula according to financial principles. For example an incorrect formula for a discounted cash flow.

3. Logical errors in Excel. This category includes incorrect application of a formula function. For example an IRR (Internal Rate of Return) function is used instead of an XIRR (Internal Rate of Return for a Series of Cash Flows) function.

4. Interface errors. This category contains incorrect or incomplete references to external sources, other spreadsheets or Pivot tables that are not up to date.

5. Input errors. Typing errors and incorrect assumptions are included here.

6. User related errors. This category contains the incorrect use of copied values and formulas (instead of correct references) or incorrect use of filters and sorting.

7. Control environment errors. These kinds of errors are caused by a lack of controls within the spreadsheet. For example formulas that are accidentally overwritten by fixed numbers, unauthorized changes and the use of wrong versions.

## 2.2 Overview of spreadsheet tools

The usage of spreadsheets is widespread [Hermans, 2012], because they are flexible and easy to use. Hermans [2012] states that half of all spreadsheets that a company has are used as a basis for decision making. Spreadsheets are also very commonly used for financial reporting purposes. Those spreadsheets have to be audited by an auditor. The extent to which a spreadsheet will be audited substantively depends on several factors. For example if the client maintains an effective internal control system to keep the integrity of a spreadsheet at the desired level, this results in lesser substantive testing by the auditor. A spreadsheet representing a material account in the financial statements results typically in more substantive testing.

Auditors should take into consideration that there is an increased likelihood of a misstatement in information provided in the form of spreadsheets. If an error causes a material misstatement and the auditor does not detect this error before certifying the financial statements, the impact on the image and the financial position of the auditor can be huge when this becomes public. The impact of the increased likelihood of misstatements in spreadsheet on the audit of spreadsheet in the context of legislation is outside the scope of this research.

Various organizations have developed spreadsheet tools that support error correction in spreadsheets. A list of spreadsheet auditing tools can be found at Resources for Spreadsheet Analytics, https://sites.google.com/a/usfca.edu/business-analytics/development-management/checking-auditing.

Research on spreadsheet tools is rather limited and conclusions are contradictory. Abraham and Erwig [2007] tested the spreadsheet tool Ucheck. In this research, spreadsheets created by university students were used to examine the effectiveness of Ucheck. These spreadsheets were evaluated by high school teachers. The researchers found that Ucheck does support users in correcting unit errors. Unit errors are categorized as input errors in our taxonomy. Since this research did not examine spreadsheets created by companies, it provides limited evidence that this spreadsheet tool will be useful for auditors when auditing spreadsheets. Powell, Baker and Lawson [2008a] developed an auditing protocol to find errors in operational spreadsheets. They tested this protocol together with spreadsheet auditing tools XL Analyst and Spreadsheet Professional amongst current undergraduate and graduate students in business or engineering and recent alumni of these programs. They found that the auditing software generated a high percentage of false positives and false negatives. However in their believe the use of auditing software is far more effective in identifying errors than unassisted auditing. They also found that auditors developed skills that allowed them to understand the formal structure of a complex spreadsheet developed a sense of where errors were likely to occur. Other spreadsheet auditing tools such as Spreadsheet Detective, Excel Auditor and Operis Analysis Kit were also subjected to research [Anderson, 2004]. The researchers concluded that these spreadsheet tools were very effective in detecting mechanical errors. Mechanical errors are according to Panko and Halverson [1996] errors arising from typing or pointing errors, so in our research we classify them as reference and input errors. In the study of [Anderson, 2004] the spreadsheet tools detected values stored as text in 82% of the cases and incomplete ranges were detected in 55% of the cases. Despite these good rates of error detection, the tools were unable to correctly flag errors in logic like the omission of a variable or an operator precedence error. These errors were caught in 18% and 9% of the cases respectively. In our taxonomy these errors are likely the cause of cells containing an incorrect formula according to financial principles or logical errors in Excel.

The spreadsheet tools described previously focused on static analysis of spreadsheets. The spreadsheet tools discussed next provide broader functionalities, including risk analysis and visualization. Hermans [2012] concludes that tools containing risk analysis and visualization functionalities contribute to a more effective and efficient spreadsheet audit. Panko and Aurigemma [2010] find that two kinds of inspection auditing software (Excel Error Check and Spreadsheet Professional), which include these functionalities, were almost useless for correctly flagging natural human errors in spreadsheets. The human inspectors found 54 errors from the total of 97 errors, as opposed to 5 errors that were flagged by the spreadsheet tools.

Microsoft itself came up with an add-in in Microsoft Office Excel 2013, to better comprehend the issues of auditing spreadsheets. This add-in is called Spreadsheet Inquire and offers various improvements compared to earlier versions of Microsoft Office Excel. These improvements are along the lines of the needs specified by Hermans [2012], as described above. O'Beirne [2013] examined this add-in and states that the functionalities of Spreadsheet Inquire are undeveloped compared to current spreadsheet tools. He found that visualization of relations between sheets from complex spreadsheets is not possible. Risk analysis is also not possible in Excel Spreadsheet Inquire. So the improvements that the add-in Spreadsheet Inquire should offer do not seem to be helpful compared to current spreadsheet tools.

Since the previous spreadsheet tools in our opinion do not provide enough support for auditors in auditing spreadsheets, we are looking at a different spreadsheet tool called PerfectXL. Hermans is the founder of Infotron, which developed this spreadsheet tool according to the research she did between 2008 and 2012. This spreadsheet tool should support auditors by providing a risk analysis function and a visualization function. Hermans [2012] determined several situations that proved high risk for causing errors in spreadsheets. Risk analysis should therefore help the auditor to focus on calculations in spreadsheets that have a high risk of containing errors. The risk analysis functionality highlights risky areas in the spreadsheet as is shown in figure 1.

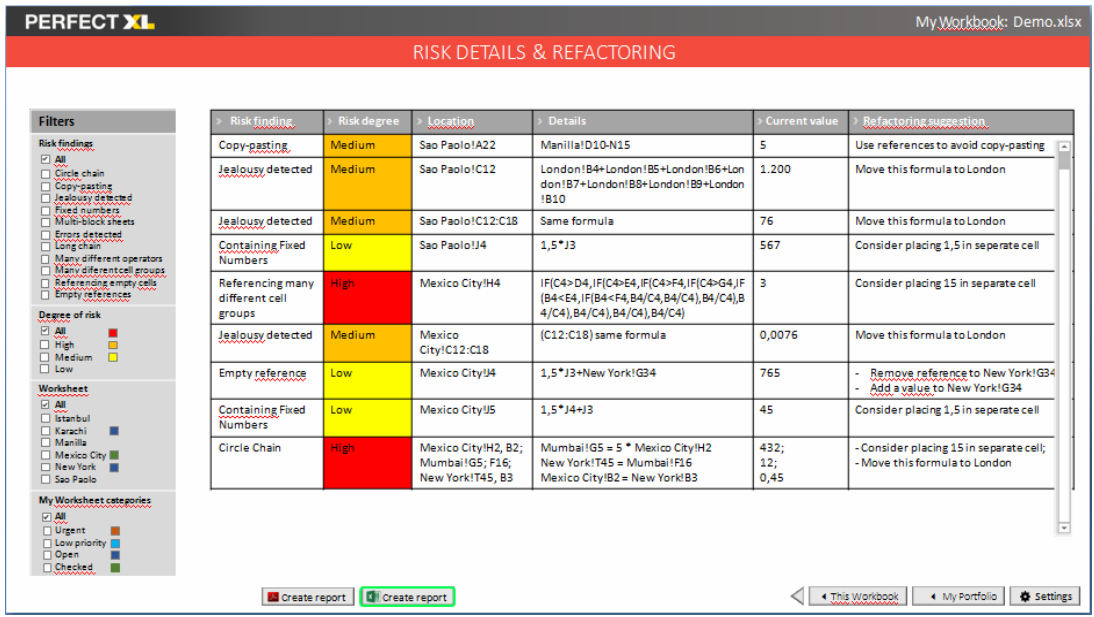

Figure 1: Risk analysis and refactoring tips as an overview for a whole spreadsheet.

A visualization functionality should also help the auditor to understand relations between different sheets of a spreadsheet. The visualization functionality in PerfectXL is available on a spreadsheet level basis which shows dependencies between sheets. Figure 2 shows the visualization of a workbook. The blocks represent the worksheets within the workbook. The external sources (worksheets from other workbooks in Excel) linked to a sheet are recognizable by the orange colour. Hidden and very hidden sheets are presented as relatively light blue and grey blocks.Thick arrows between the blocks indicate large dependencies.

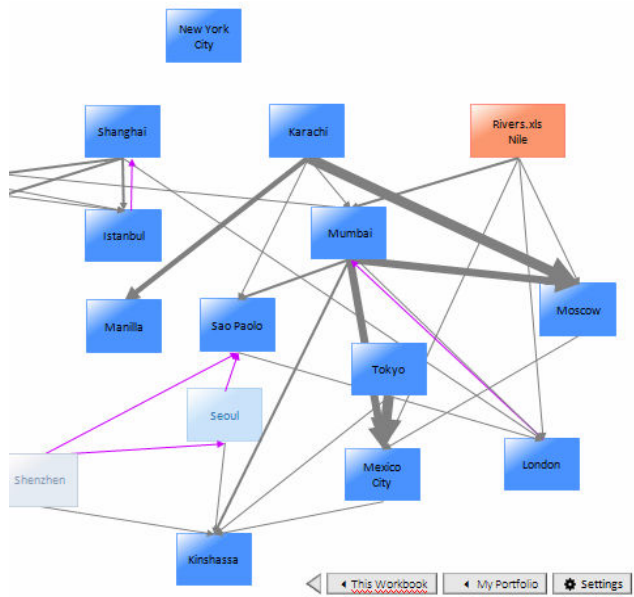

Figure 2: Visualization of a part of a spreadsheet where external sources and hidden sheets are made visible.

The visualization functionality is also available on a sheet level basis (figure 3) whereas the content of the sheet is divided into different categories. Orange cells are labelled as text, yellow cells are labelled as singe numbers, blue cells indicate that a formula is used. The purple lined boxes show the range of consistent formulas.

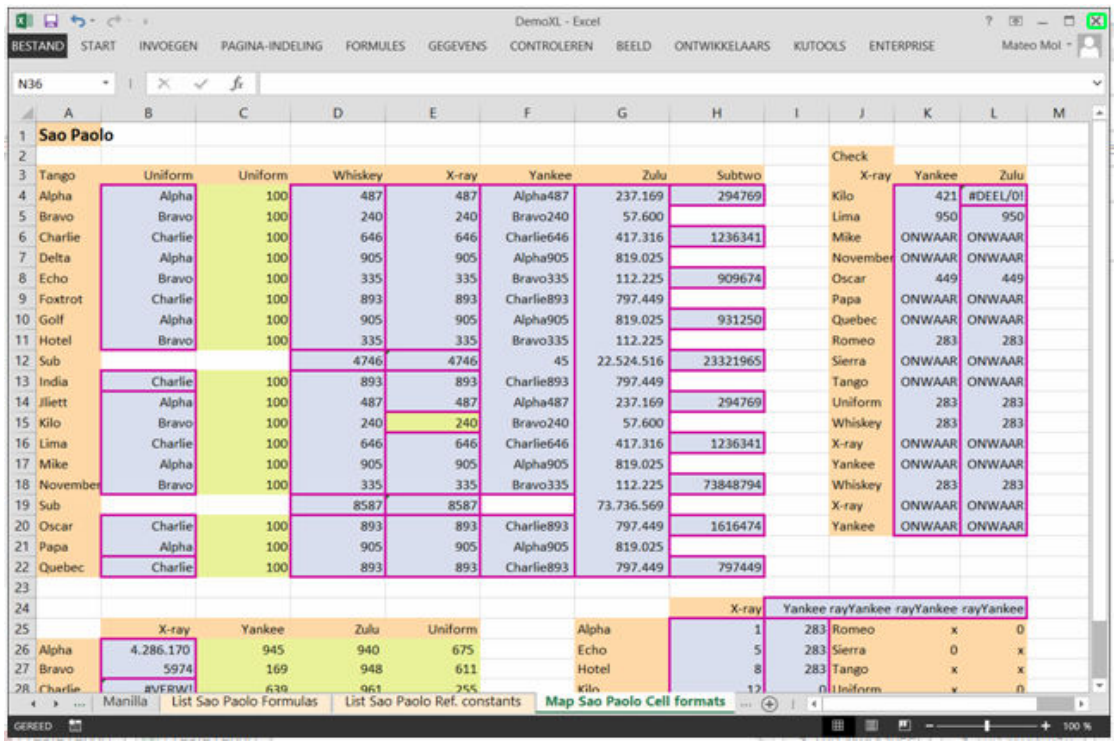

Figure 3: Visualization of the content of a sheet including specifications of cells.

A full overview and explanation of PerfectXL is presented in appendix 1. A free trial of this tool is available online at www.infotron.nl.

**3 METHODOLOGY**

**3.1 Research questions**

This research aims to answer the following questions:

> **R1** Are common errors recognized by auditors?
>
> **R2** To what extent does PerfectXL contribute to a more efficient and effective audit?
>
> **R3** Which functionalities should PerfectXL possess in order to contribute towards a more efficient and effective audit?

**3.2 Research design**

This research is designed to determine the perceptions of auditors on the usefulness of PerfectXL. 8 auditors, all working for the same big four audit firm in the Netherlands, participated in this research. The auditors who were selected have a working experience of over four years. The auditors were interviewed in a semi-structured setting. The interviews were performed over a period of four weeks, with two interviews each week. Each interview was recorded and transcribed verbatim. The results obtained in the analysis of these interviews were adapted in the questions for later interviews. A full overview of questions asked is included in appendix 2. Normally, the interview started with general questions about the length of service at the audit firm and the customers the auditors are working for. Then there were some more specific questions about common mistakes in Excel made by clients. The spreadsheet tool PerfectXL was then demonstrated to the auditors. Finally, the auditors were asked to give their opinion according to positive, missing and negative aspects of the tool.

The interviews with the auditors were semi-structured because this gave the auditors the ability to voice their opinions without being tied in to answering only structured questions. The tenure of work experience was set to ensure that the auditors had obtained enough experience auditing spreadsheets. Since the PerfectXL tool was not ready at the time the research took place, a presentation of the future tool of PerfectXL was given to the auditors to show the capabilities of PerfectXL. The sheets presented the functionalities visualization and risk analysis and gave an impression of the interface of PerfectXL. For an explanation of these functionalities, refer to section 2.2. Each of those sheets was carefully explained to the auditor by the interviewer, who had significant knowledge of the future tool of PerfectXL. A brief overview of this presentation is included in appendix 2. A full overview of this presentation is available online at figshare.

## 4 RESULTS

### 4.1 Most common errors

The auditors agree with the occurrence of errors in spreadsheets. The table below shows the scores from the interviews of the most common errors derived from literature.

| **Most common errors** | **Yes** | **No** | **No answer** |
|---|---|---|---|
| *Incorrect range of formulas* | 75% | 12.5% | 12.5% |
| *Typing errors* | 12.5% | 12.5% | 75% |
| *External sources not up to date* | 12.5% | 12.5% | 62.5% |
| *Copied formulas and values that lead to errors* | 50% | 37.5% | 12.5% |
| *Incorrect negative numbers* | 62.5% | 25% | 12.5% |
| *Incorrect input* | 12.5% | 12.5% | 75% |
| *Unfixed error message from Excel* | 12.5% | 12.5% | 75% |

Table 1: Most common errors noted by auditors from the sample

Most of the common errors that are recognized by auditors are incorrect range of formulas, copied formulas and values that lead to errors and incorrect negative numbers. From the interviews, there is no evidence that typing errors, out of date external sources, incorrect input or unfixed error messages from Excel are recognized common errors from an auditor's perspective. This can be concluded because the auditors did not mention these errors when asked which errors they frequently encountered in spreadsheets prepared by the client. The last one – unfixed error messages from Excel – seems understandable because an error message from Excel is easily recognized by the client. It is therefore plausible that the client corrects these mistakes before he provides the spreadsheet to the auditor.

**4.2 Perceptions of functionalities of PerfectXL**

The table below shows the answers of auditors on the basis of the demonstration of the demo tool.

| **Opinion about the tool** | **Positive** | **Doubtful** | **Negative** | **No answer** |
|---|---|---|---|---|
| *Risk analysis* | 37.5% | 50% | 12.5% | 0% |
| *Visualization of spreadsheet logic in general* | 62.5% | 12.5% | 0% | 25% |
| *Analysis on spreadsheet level* | 50% | 0% | 12.5% | 37.5% |
| *Analysis on sheet level* | 62.5% | 0% | 0% | 37.5% |
| | | | | |
| | **Agree** | **Doubtful** | **Disagree** | **No answer** |
| *The tool is indicating direction for the audit* | 62.5% | 0% | 0% | 37.5% |
| *The tool is efficient and effective* | 0% | 25% | 0% | 75% |

Table 2: The perceptions of the auditors from the sample about the usefulness of a spreadsheet tool

The attitude of auditors towards the different functionalities of a spreadsheet tool was mixed. Most auditors were favorable towards the visualization. The analysis on sheet level obtained the most favorable results. The auditors found this functionality helpful for quickly checking for the internal consistency of formulas. Some auditors thought that the analysis on spreadsheet level would help them auditing spreadsheets by better understanding the spreadsheet logic. The auditors were doubtful about the risk analysis.

The last questions were about the attitude in general towards a spreadsheet tool. Most auditors found that the tool indicated a direction for the audit. This could be dangerous if an auditor trusts the tool and overlooks further analysis if no risks are indicated. One respondent formulated this as follows:

> "What you saw in the visualization with the colors, green, orange and red. That is very nice to have as a guideline. When you see a sheet that is green, then it will be correct. If it is red then it is time to have a look at what precisely is going on."

A last observation is that two of the auditors specifically mentioned that they were doubtful that the spreadsheet tool would improve the efficiency and effectiveness of the audit. The following quote indicates that concern:

> "What I find difficult is that you could lose yourself in such analysis. So the tool provides various issues and you could check every formula but I am doubtful how 1 could use this efficiently and effectively in my audit."

# 5 CONCLUSION

The goal of this research was to determine the perceptions of the usefulness of PerfectXL in auditing spreadsheets. We defined these perceptions by answering three research questions. Most of the errors defined as common by previous researchers were recognized by the auditors. This indicates the importance of an effective audit by using a spreadsheet tool. Furthermore the auditors agreed that the visualization functionality supported a more effective audit. Especially the visualization on a sheet level basis as referred to in Section 2.2 provided high added value, in the opinion of the auditor. Despite these positive opinions from the auditors, some specifically mentioned that they were doubtful about the effectiveness and efficiency of using PerfectXL in their audit of spreadsheets. We do however believe that the majority of the auditors would appreciate using PerfectXL in their audit of spreadsheets.

More specifically, we give the following answers regarding the research questions.

> **R1** Are common errors recognized by auditors?

Four out of seven scientifically defined common errors are not recognized by the auditors. For the error messages from Excel, there could be an explanation from the auditor's perspective because these errors are so obvious that they could easily be detected and corrected by the client before providing the spreadsheet to the auditor.

> **R2** To what extent do spreadsheet tools contribute to a more efficient and effective audit?

The results for this research question are broad because of the diversity of opinions on the spreadsheet tool from the respondents. The results do indicate, however, that some tool functionalities are helpful in the audit. The functionalities that we refer to in this research are visualization and risk analysis.

> **R3** Which functionalities should spreadsheet tools have in order to contribute to a more efficient and effective audit?

The results clearly indicate that the majority of the auditors had the perception that a visualization is useful in order to perform a more efficient and effective audit. There is doubt about the added value of a risk analysis. The respondents indicate that is not obvious that risk analysis results in a more efficient and effective audit.

Furthermore, this research sheds light on the usefulness of other spreadsheet tools, in addition to PerfectXL. Because the functionalities of PerfectXL are scientifically composed and more spreadsheet tools offer the same functionalities, a similar experiment could also apply to these spreadsheet tools. This research is also unique in examining the needs of auditors in auditing spreadsheets with spreadsheet tools.

Further research needs to be done in order to determine the improvement of efficiency and effectiveness of audits through the use of spreadsheet tools within the audit. The quantitative effect of using a spreadsheet tool could be determined by performing an experiment. This experiment could be done with spreadsheet tool PerfectXL, because this tool provides functionalities needed by auditors, according to Hermans [2012]. The research by Aurigemma and Panko [2010] provides a good example for the research design of this experiment. The spreadsheets used to perform this experiment could be randomly chosen from spreadsheets provided by clients. As we know from the research by Aurigemma and Panko [2010], some errors are not correctly flagged by spreadsheet tools. The aim of this experiment should not be to look at correctly flagging errors by PerfectXL but should rather focus on differences in effectiveness and efficiency between groups of auditors who audit a spreadsheet with a tool and without a tool. This experiment thus has to measure the percentage of errors detected and the time it took to audit the spreadsheet. These measures, compared between the groups, would give an indication of the increase in effectiveness and efficiency through the use of a spreadsheet tool when auditing spreadsheets.

## ACKNOWLEDGEMENT

The authors thank the HU Utrecht University of Applied Sciences/FAI and Delft University of Technology for their support in setting up this combined student research project.

## 6 REFERENCES

Anderson, W. (2004). A *Comparison of Automated and Manual Spreadsheet Detection.* (Master's thesis, Massey University, Albany, New Zealand.)

Abraham, R., & Erwig, M. (2007). Ucheck: A spreadsheet type checker for end users. *Journal of Visual Languages and Computing, vol. 18*, pp. 71–95.

Aurigemma, S., & Panko, R. R. (2010). The detection of human spreadsheet errors by humans versus inspection (auditing) software. *arXiv preprint arXiv:1009.2785*.

de Ruijter, I., & Potjer, J. (2006). Integriteit van spreadsheets. *MCA Tijdschrift voor Organisaties in Control*, pp. 4-10.

*EuSpRIG Horror Stories*. (2012). Retrieved from EuSpRIG: http://eusprig.org/horror-stories.htm

Hermans, F. F. J. (2012). *Analyzing and visualizing Spreadsheets* (Doctoral dissertation, PhD thesis, Software Engineering Research Group, Delft University of Technology, Netherlands).

Kreie, J., Cronan, T. P., Pendley, J., & Renwick, J. S. (2000). Applications development by end-users: can quality be improved?. *Decision support systems*, *29*(2), pp. 143-152.

O'Beirne, P. (2014). Excel 2013 Spreadsheet Inquire. *arXiv preprint arXiv:1401.7586*.

Olson, J. R., & Nilsen, E. (1987). Analysis of the cognition involved in spreadsheet software interaction. *Human-Computer Interaction*, *3*(4), pp. 309-349.

Panko, R. R. (1998). What we know about spreadsheet errors. *Journal of Organizational and End User Computing (JOEUC)*, *10*(2), pp. 15-21.

Panko, R. R. (2008). Spreadsheet errors: What we know. what we think we can do. *arXiv preprint arXiv:0802.3457*.

Panko, R. R., & Aurigemma, S. (2010). Revising the Panko–Halverson taxonomy of spreadsheet errors. *Decision Support Systems*, *49*(2), 235-244.

Panko, R. R., & Halverson Jr, R. P. (1996). Spreadsheets on trial: a survey of research on spreadsheet risks. *System Sciences, 1996., Proceedings of the Twenty-Ninth Hawaii International Conference on,* (Vol. 2, pp. 326-335). IEEE.

Powell, S. G., Baker, K. R., & Lawson, B. (2008a). An auditing protocol for spreadsheet models. *Information & Management*, *45*(5), 312-320.

Powell, S. G., Baker, K. R., & Lawson, B. (2008b). A critical review of the literature on spreadsheet errors. *Decision Support Systems*, *46*(1), 128-138.

Powell, S. G., Baker, K. R., & Lawson, B. (2009a). Errors in operational spreadsheets. *Journal of Organizational and End User Computing (JOEUC)*,*21*(3), 24-36.

Powell, S. G., Baker, K. R., & Lawson, B. (2009b). Impact of errors in operational spreadsheets. *Decision Support Systems*, *47*(2), 126-132.

<>
Proceedings of the EuSpRIG 2015 Conference "Spreadsheet Risk Management" ISBN: 978-1-905404-52-0
Copyright © 2015, European Spreadsheet Risks Interest Group (www.eusprig.org) & the Author(s)
Page 11/15

## APPENDIX 1: BRIEF OVERVIEW OF PERFECTXL

The images below are fragments of the presentation that was shown to the auditors to give an overview of the design and functionalities of PerfectXL:

Figure 4: Risk analysis and refactoring tips as an overview for the whole spreadsheet.

The image above is an overview of the risk analysis of PerfectXL. The tool highlights the following risks:

- fixed numbers in formulas;
- unusual ranges which are detected through inconsistency in a column of consistent formulas;
- formulas that contain a large number of references to another sheet;
- mutiple functions in one formula;
- many cell references in one formula;
- a long chain of formulas;
- copied formulas;
- references to empty cells;
- error messages from Excel.

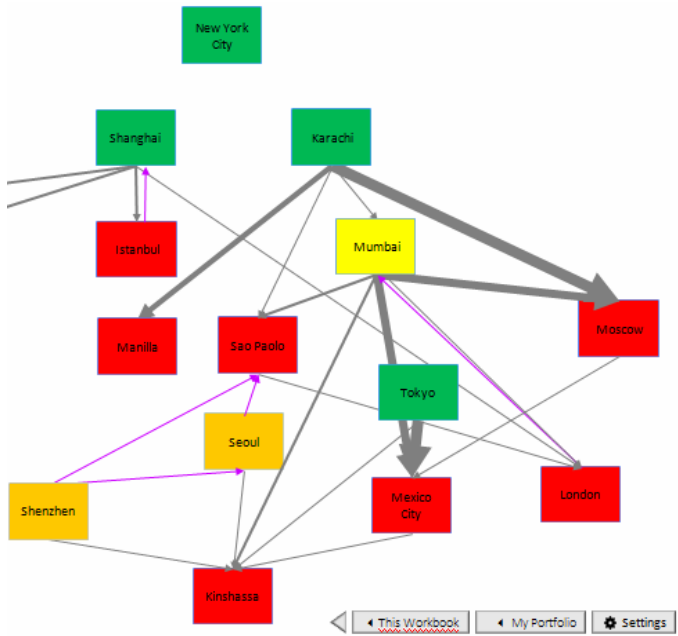

Figure 5: Visualisation with an overview of the degree of risks in the different sheets.

The image above represents the structure of a part of sheets in a spreadsheet. The colours reflect the degree of risk in the separate sheets. A grey arrow shows that there is a link between two sheets. The colour grey stands for a reference to a cell from a sheet on the right hand side of the sheet from which the grey arrow originates. The opposite is the case for a purple arrow.

Figure 6: Visualization of a sheet using risk analysis.

This figure shows the risks that come out of the risk analysis. The results are now presented in the sheet. Yellow coloured cells have a low risk of containing an error.

Orange coloured cells have a moderate risk of containing an error. Red coloured cells have a high risk of containing an error.

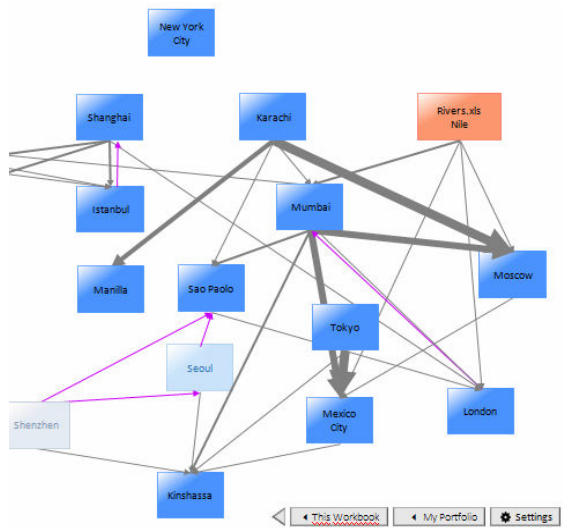

Figure 7: Visualization of the whole spreadsheet where external sources and hidden sheets are made visible.

The above figure shows the same arrows as in figure 4. The link between external sources and sheets can be seen in this figure. These external sources are recognizable by the orange colour. Hidden and very hidden sheets are presented as relatively light blue and grey blocks.

Figure 8: Visualization of the content of a sheet including specifications of cells.

In the above figure, orange cells are labelled as text, yellow cells are labelled as singe numbers, blue cells indicate that a formula is used. The purple lined boxes show the range of consistent formulas.

## APPENDIX 2: FIXED QUESTIONS IN THE INTERVIEWS

**General questions:**

- How long have you been employed at the audit firm?
- What type of clients do you have?

**Questions relating to the use of Excel by clients and auditors and the problems they have while using Excel:**

- At what type of customers do you come across spreadsheets most often?
- For what specifications do they use spreadsheets?
- How do you audit spreadsheets? Do you use tools for the audit of spreadsheets? Do you miss anything for this?
- What errors do you detect in spreadsheets?
- Do you sometimes have problems with understanding/auditing spreadsheets?
- Which of the following situations do you see in spreadsheets and how often do you see these situations?
  - Fixed numbers in formulas
  - Copied values and formulas
  - Negative numbers
  - Hidden cells
  - Overly long formulas
  - Incorrect ranges

**Questions relating to the tool:**

- What do you like about the tool?
- What would you like to see more in the tool?
- What do you think is bad/unnecessary in the tool?